\documentclass[12pt]{iopart}

\usepackage{rotating}  
\usepackage[font=small,labelfont=bf]{caption}
\begin{document}

\title[Spectroscopy of Light Baryons: $\Delta$ Resonances]{Spectroscopy of Light Baryons: $\Delta$ Resonances }

\author{Chandni Menapara and Ajay Kumar Rai}

\address{Department of Physics, Sardar Vallabhbhai National Institute of Technology, Surat-395007, Gujarat, India}
\ead{chandni.menapara@gmail.com}
\vspace{10pt}
\begin{indented}
\item[]
\end{indented}

\begin{abstract}
The spectroscopy of light, strange baryons have been an important aspect with still unknown resonances and intrinsic baryonic properties. The present document is focused on the $\Delta$ baryons unlike earlier work, here all the four isospin states have been separately obtained treating u and d quarks with different constituent masses. The theoretical framework for calculating the resonance masses is hypercentral Constituent Quark Model (hCQM). The confining potential is taken as of linear form alongwith the addition of spin-dependent as well as first order correction term. The results obtained have been compared with many varied approaches as well as experimental available states. In addition to mass spectra, Regge trajectories for (n,$M^{2}$) and  (J,$M^{2}$) have been plotted. The magnetic moment as well as transition magnetic moment have been calculated using effective quark mass. The radiative decay width has been studied for $\Delta^{+}$ and $\Delta^{0}$ channels. 
\end{abstract}

%
\vspace{2pc}

\section{Introduction}

The quest towards the understanding of internal dynamics of hadrons has been the driving force to hadron spectroscopy. It is a prime means for our knowledge of QCD in the low-energy regime \cite{crede}. The search for the missing resonances has lead us to explore various theoretical and phenomenological approaches to obtain the excited baryon spectrum and to observe them has been central goal of experimental facilities. The experiments at Jefferson Lab, MAMI, ELSA \cite{elsa}, GRAAL, HADES-GSI \cite{hades} have been striving through years to collect as much information about excited state of hadrons as possible. The upcoming facilities PANDA at FAIR-GSI shall be dedicated to light, strange baryons \cite{panda1,panda2,panda3}. \\\\
The first and foremost members of octet and decuplet families, nucleon N and delta $\Delta$ baryon have always been of interest. Various decay of strange as well as other heavy baryons are ultimately reaching to these light baryons. The isospin partners u and d quarks have generally been treated at the same footings however, it leads to 2 and 4 respective isospin states for N and $\Delta$. This study is motivated by our attempt to study P and N separately and now extending to all the four isospins of $\Delta$ \cite{epjweb}. The possible four combinations of the symmetric wave function gives four $\Delta$ particle with isospin $I = \frac{3}{2}$ as: \\
\begin{center}
\begin{tabular}{c}
$\Delta^{++}$ (uuu, $I_{3} = \frac{3}{2}$)\\
$\Delta^{+}$ (uud, $I_{3} = \frac{1}{2}$) \\
$\Delta^{0}$ (udd, $I_{3} = -\frac{1}{2}$) \\
$\Delta^{-}$ (ddd, $I_{3} = -\frac{3}{2}$)
\end{tabular}
\end{center}
Over the years, $\Delta$s have been observed through pion-nucleon decays \cite{barnicha, pedroni} and photoproduction decays. The study of $\Delta$ is not only limited to high energy realm but also in the field of astrophysics wherein $\Delta$s are observed in neutron stars as a puzzle too \cite{sahoo}. Historically, the presence of $\Delta$(1232) has been an essential step towards color degree of freedom. Incorporating the recent additions, 8 four star, 4 three star and many other experimental status have been explored with the values ranging from $J=\frac{1}{2}$ to $J=\frac{15}{2}$ and still many states are awaited of confirmation of existence \cite{pdg}. Not many baryons are known with such a number of states. Also, $\Delta$ being the lightest member with the presence of electric quadrupole moment makes it interesting to dig deep into the shape and structure of the baryon \cite{ramalho}. Also, the MicroBooNE collaboration has recently reported $\Delta$(1232) radiative decay through neutrino induced neutral current \cite{microboone}. \\\\ 
Other than experiments, $\Delta$ resonances have been focused through various theoretical and phenomenological approaches. To name a few, Isgur-Karl model \cite{isgur} with modified relativised approach \cite{capstick}, spontaneous chiral symmetry breaking through Goldstone-boson exchange\cite{papp, zahra}, semi-relativistic model \cite{rajabi}, QCD SUM Rules  \cite{azizi}, light-front model \cite{vary} and light-front relativistic \cite{aznauryan}, Lattice QCD \cite{edwards} and covariant Faddeev approach \cite{sanchis}, relativistic quark with instanton-induced forces \cite{loring} and many more. Others are also described in section 3 wherein used for comparison of the calculated data.\\\\
The paper is organized as follows: the theoretical framework describes the non-relativistic hypercentral Contituent Quark Model (hCQM) which has been employed for the present study. The results for radial and orbital resonance masses are tabulated alongwith the experimental known states as well as other theoretical and phenomenological models ranging from S to F-wave for all four $\Delta$ partners. Later sections 4, 5, and 6 focus on Regge trajectories, magnetic moments and radiative decay widths for the calculated masses. Finally, the article is concluded with the implications drawn from the study. 

\section{Theoretical Formalism}
\label{sec-1}
In a constituent quark model, the baryons are considered as system of three constituent quarks, wherein all the effects are parametrized in the form of larger constituent quark mass compared to that of a valence quark. These constituent quarks are described as valence quarks dressed with gluons and quark-antiquark pairs and all the other interaction leading to the total baryon mass. Here one such constituent quark model with hypercentral and non-relativistic formalism has been employed. In our earlier work, the constituent quark mass for u and d quarks were taken to be similar \cite{zalak19,cpc1,cpc2,aip,dae,keval20,keval18,zalak18}. In the relativistic model for the study of strange hyperons through quark-diquark, u and d are considered with similar constituent quark mass \cite{galkin}. However, in the present work, the u and d constituent quarks masses have been modified respectively as $m_{u}=290$ and $m_{d}=300$ which allowed us to segregate the four isospin states of $\Delta$ baryon. \\
The three body interaction of quarks inside a baryon is described in the form of Jacobi coordinates $\rho$ and $\lambda$ which are obtained based on inter-quark distance $r_{i}$.\\
\begin{equation}
{\bf \rho} = \frac{1}{\sqrt{2}}({\bf r_{1}} -{\bf r_{2}}); \; \; {\bf \lambda} = \frac{1}{\sqrt{6}} ( {\bf r_{1}} + {\bf r_{2}} - 2{\bf r_{3}})
\end{equation}
The hypercentral Constituent Quark Model (hCQM) is reached through hyperradius x and hyperangle $\xi$. 
\begin{equation}
x = \sqrt{{\bf \rho^{2}} + {\bf \lambda^{2}}}
; \; \; \xi = arctan(\frac{\rho}{\lambda})
\end{equation}
The potential to account for confinement and asymptotic freedom of quarks within a baryon is taken to be Coulomb-like part and a linear term as a confining part. As the model itself suggests, the potential is solely depended on the hyperradius x. It is noteworthy here that x indirectly is being contributed with the three-body interaction.  
\begin{equation}
 V(x) = -\frac{\tau}{x} + \alpha x
\end{equation}
To take into account the possible angular momentum quantum number J, spin-dependent terms are also added to the earlier potential terms. 
\begin{equation}
V_{SD}(x) = V_{SS}(x)({\bf S_{\rho}\cdot S_{\lambda}}) +  V_{\gamma S}(x)({\bf \gamma \cdot S}) + V_{T}\times [S^{2}- \frac{3({\bf S\cdot x})({\bf S\cdot x})}{x^{2}}]
\end{equation}
Here, $V_{SS}(x)$, $V_{\gamma S}(x)$ and $V_{T}(x)$ are spin-spin, spin-orbit and tensor terms respectively \cite{voloshin}.
\begin{equation}
V_{SS}(x)=\frac{1}{3m_{\rho}m_{\lambda}}\nabla^{2}V_{V} 
\end{equation}

\begin{equation}
V_{\gamma S}(x)==\frac{1}{2m_{\rho}m_{\lambda}x}(3\frac{dV_{V}}{dx}-\frac{dV_{S}}{dx})
\end{equation}

\begin{equation}
V_{T}(x)=\frac{1}{6m_{\rho}m_{\lambda}}(3\frac{d^{2}V_{V}}{dx^{2}}-\frac{1}{x}\frac{dV_{V}}{dx})
\end{equation}
where $V_{V}$ and $V_{S}$ are the vector and scalar part of potential.
In addition to above terms, a first order correction term with $\frac{1}{m}$ dependence has also been incorporated. 
\begin{equation}
V^{1}(x)= -C_{F}C_{A}\frac{\alpha_{s}^{2}}{4x^{2}}
\end{equation}
where $C_{F}$ and $C_{A}$ are Casimir elements of fundamental and adjoint representation. \\
Thus the  hyper-radial part of the wave-function as determined by hypercentral Schrodinger equation is \cite{giannini, gianinni1}
\begin{equation}
\left[\frac{d^{2}}{dx^{2}} + \frac{5}{x}\frac{d}{dx} - \frac{\gamma(\gamma +4)}{x^{2}}\right]\psi(x) = -2m[E-V(x)]\psi(x)
\end{equation}
Here $\gamma$ replaces the angular momentum quantum number by the relation as $l(l+1) \rightarrow \frac{15}{4} + \gamma (\gamma + 4)$.
The final Hamiltonian is as follows:
\begin{equation}
H = \frac{P^{2}}{2m} + V(x)  + V_{SD}(x) + V^{1}(x)
\end{equation}
The Schrodinger equation with the hyper-radial part is numerically solved for calculating the excited state masses \cite{lucha}. Details of the model can be found in articles \cite{cpc1,zalak18}

\section{Results and Discussions}
Using the above potential model, the masses are computed for 1S-5S, 1P-3P, 1D-2D, 1F states including few states from 1G, 1H and 1I which were not obtained earlier. The excited states are recalculated for all these isospin states of $\Delta$ baryon and compared with various results as shown in the table below. \\\\
The models for comparison vary in wide range starting from some early works to recent ones. Recently all light and strange baryons have been studied through Bethe Ansatz method with U(7) by an algebraic method \cite{amiri}. The earlier algebraic method has been discussed by R. Bijker et.al. to study baryon resonances in terms of string-like model \cite{bijker}. A. V. Anisovich et. al. has reproduced the N and $\Delta$ spectrum using the multichannel partial wave analysis of pion and photo-induced reactions \cite{thoma}. The quark-diquark model using Gursey Radicati-inspired exchange interaction has been studied \cite{s5, s15}. The semi-relativistic constituent quark model, classification number describing baryon mass range \cite{chen}, mass formula obtained by Klempt \cite{klempt}, dynamical chirally improved quarks by BGR \cite{bgr} are among various models. \\\\
As the present work has attempted to separate the isospin states of all four $\Delta$ baryons, the ground state masses nearly vary by 2 MeV for each one but within the PDG range. A similar trend is observed in higher radial excited states of S-wave. The 2S mass predicted is very much near to the algebraic model \cite{amiri}. \\\\
The negative parity states have states ranging from 4 to 1 star status by PDG. It is observed from the table that with increase in J value, the predicted masses are under-predicted. However, not many models have different masses for every spin-parity state. In case of F-wave, $J=\frac{7}{2}$ state is the only known by PDG. The present masses are nearly 100 MeV below the range and hardly any comparison is obtained. \\\\

\begin{small}
{\bf Table 1}: Resonance Masses of $\Delta$ Baryons and comparison of various model (in MeV)\\
\rotatebox{90}{
\setlength{\tabcolsep}{1pt}
\begin{tabular*}{240mm}{@{\extracolsep{\fill}}cccccccccccccccccccc}
\hline
 State &  $J^{P}$ & $\Delta^{++}$ & $\Delta^{+}$ & $\Delta^{0}$ & $\Delta^{-}$ &\cite{cpc1} & PDG\cite{pdg} & Status & \cite{amiri} &  \cite{zahra} & \cite{s5} & \cite{s15} & \cite{rajabi} & \cite{gianinni1} & \cite{klempt} & \cite{isgur} & \cite{capstick} & \cite{chen} & \cite{bgr}\\
\hline
1S & $\frac{3}{2}^{+}$ & 1228 & 1230 & 1232 & 1235 & 1232 & 1230-1234 & **** & 1245 & 1232 & 1235 & 1247 & 1231 & 1232 & 1232 & 1232 & 1230 & 1232 & 1344 $\pm$ 27\\ 
2S & $\frac{3}{2}^{+}$ & 1603 & 1606 & 1610 & 1615 & 1611 & 1500-1640 & **** & 1609 &  1659.1 & 1714 & 1689 & 1658 & 1727 & 1625 &  & & 1600\\
3S & $\frac{3}{2}^{+}$ & 1922 & 1926 & 1932 & 1941 &
1934 & 1870-1970 & *** & & 2090.2 & 1930 & 2042 & 1914 & 1921 & 1935 & & & 1920\\
4S & $\frac{3}{2}^{+}$ &  2241 &	2248 &	2257 &	2270 &
 2256 & - & - & \\
5S & $\frac{3}{2}^{+}$ &  2559	& 2570 &	2584 &	2602 &
 2579 & - & - & \\
\hline
1P & $\frac{1}{2}^{-}$ & 1618 & 1625 & 1630	& 1634 & 
 1625 & 1590-1630 & **** & 1711 & 1667.2 & 1673 & 1830 & 1737 & 1573 & 1645 & 1685 & 1555 & & 1454 $\pm$ 140\\
1P & $\frac{3}{2}^{-}$ & 1585 & 1591 & 	1596 & 1603 & 
 1593 & 1690-1730 & **** & 1709 & 1667.2 & 1673 & 1830 & 1737 & 1573 & 1720 & 1685 & 1620 & & 1570 $\pm$ 67\\
1P & $\frac{5}{2}^{-}$ & 1542 & 1546 & 1552 & 1561 & 
 1550 & - & - & \\
\hline
2P & $\frac{1}{2}^{-}$ & 1943 &	1944 & 	1955 & 	1965 & 
1956 & 1840-1920 & ***  &  & & 2003 & 1910 & & 1910 & 1900 & & & & 1914 $\pm$ 322 \\
2P & $\frac{3}{2}^{-}$ & 1907 & 1911 &	1921 & 	1903 &
1919 & 1940-2060 & ** &  & &  & 1910 &  &  & 1940\\
2P & $\frac{5}{2}^{-}$ & 1859 & 1868 & 1874 & 1885 &
1871 & 1900-2000 & *** &  & & 2003 & 1910 & 1908 & & 1945 \\
\hline
3P & $\frac{1}{2}^{-}$ & 2262 & 2271 & 2280 & 2295 & 
2280 & - & * & &  &  &  &  &  &  2150\\
3P & $\frac{3}{2}^{-}$ & 2226 & 2235 & 2246 & 2260 & 
 2242 & - & - \\
3P & $\frac{5}{2}^{-}$ & 2179 & 2188 & 2199 & 2213 & 
2193 & - & - \\
\hline
1D & $\frac{1}{2}^{+}$ &  1898 & 1898 &	1911 & 	1919 &
1905 & 1850-1950 & **** & 1851 & 1873.5 & 1930 & 1827 & 1891 & 1953 & 1895 & & & 1910 & 1751 $\pm$ 190\\
1D & $\frac{3}{2}^{+}$ & 1860 & 1862 & 1873 & 1882 &
1868 & 1870-1970 & *** & 1936 & 2090.2 & 1930 & 2042 & 1914  & 1921 & 1935 & & & 1920\\
1D & $\frac{5}{2}^{+}$ & 1808 & 1814 & 1823 & 1832 &
 1818 & 1855-1910 & **** & 1934 & 1873.5 & 1930 & 2042 & 1891 & 1901 & 1895 & & & 1905\\
1D & $\frac{7}{2}^{+}$ & 1744 & 1753 & 1760 & 1771 &
1756 & 1915-1950 & **** & 1932 & 1873.5 & 1930 & 2042 & 1891 & 1955 & 1950 & & & 1950\\
\hline
2D & $\frac{1}{2}^{+}$ & 2218 & 2221 &	2234 & 	2247 &
2227 & - & - & & & & & & & & & & & 2211 $\pm$ 126 \\
2D & $\frac{3}{2}^{+}$ & 2179 & 2184 & 	2196 & 	2210 & 
2190 & - & - & & & & & & & & & & & 2204 $\pm$ 82\\
2D & $\frac{5}{2}^{+}$ & 2127 & 2135 &	2146 &	2160 &
2140 & 2015 & ** & & & & & & & 2200 \\
2D & $\frac{7}{2}^{+}$ & 2062 &	2073 & 	2084 & 	2098 & 
2078 & - & - \\
\hline
1F & $\frac{3}{2}^{-}$ & 2146 & 2153 & 2160 & 2181 &
2165 & - & - & & & & & & & & & & & 2373 $\pm$ 140\\
1F & $\frac{5}{2}^{-}$ & 2092 & 2099 & 2108 & 2126 & 
 2108 & - & - \\
1F & $\frac{7}{2}^{-}$ & 2024 & 2033 & 2043 & 2058 &
2037 & 2150-2250 & *** &  &  &  &  &  & 2200 \\
1F & $\frac{9}{2}^{-}$ & 1942 & 1953 &	1966 & 1975 & 
1952 & - & - \\
\hline
1G  & $\frac{11}{2}^{+}$ & 2132 & 2145 & 2162 & 2178 &  & 2300-2500 & **** & \\
\hline
1H & $\frac{13}{2}^{-}$ & 2326 & 2339 & 2362 & 2379 & & 2794 & ** & \\
\hline
1I & $\frac{15}{2}^{+}$ & 2512 & 2529 & 2554 & 2581 & & 2990 & ** & \\
\hline
\end{tabular*}
}
\end{small}
All the spin states for 1D are experimentally established. Here also our results for higher spin state are under-predicted compared to all other models and experiment but quite in accordance for lower J values. The resonance masses for 1G ($\frac{11}{2}$), 1H ($\frac{13}{2}$) and 1I ($\frac{15}{2}$) which were not present in previous work, have now been incorporated. However, these three states are under-predicted compared to PDG.\\\\
Even with the small difference in the isospin state masses, this study is expected to aid in the decay channel studies as in the decay of heavy baryons, the final products are the light baryons. In addition, the light strange baryons play important role in other areas including astrophysics to understand the composition of celestial bodies. With these obtained masses, we have attempted to study the Regge trajectories as well as magnetic moments of $\Delta$ baryon.

\section{Regge Trajectory}
Regge trajectories have known to be an important property to be explored in spectroscopic studies \cite{zalak16,zalak17}. The linear relation of total angular momentum quantum number J as well as principle quantum number n with the square of resonance mass $M^{2}$ is at the base. The plotting of all the natural and unnatural parity states allows us to locate if a given state is in accordance with the assigned $J^{P}$ value. 

\begin{equation}
J = aM^{2} + a_{0}
\end{equation}
 
 \begin{equation}
 n = b M^{2} + b_{0} 
\end{equation}
So, here figures 1 to 8 depicts the Regge trajectories for all the four isospin partners. It is noteworthy that all the calculated points fits well on the linear curve and are non-intersecting. This is expected to allow us to put an unknown experimental state on a given line to predict its $J^{P}$ value. 
\begin{figure}[b]

  \begin{minipage}[b]{0.49\textwidth}
  \caption{Regge trajectory $n \rightarrow M^{2}$ for $\Delta^{++}$}
    \includegraphics[width=\textwidth]{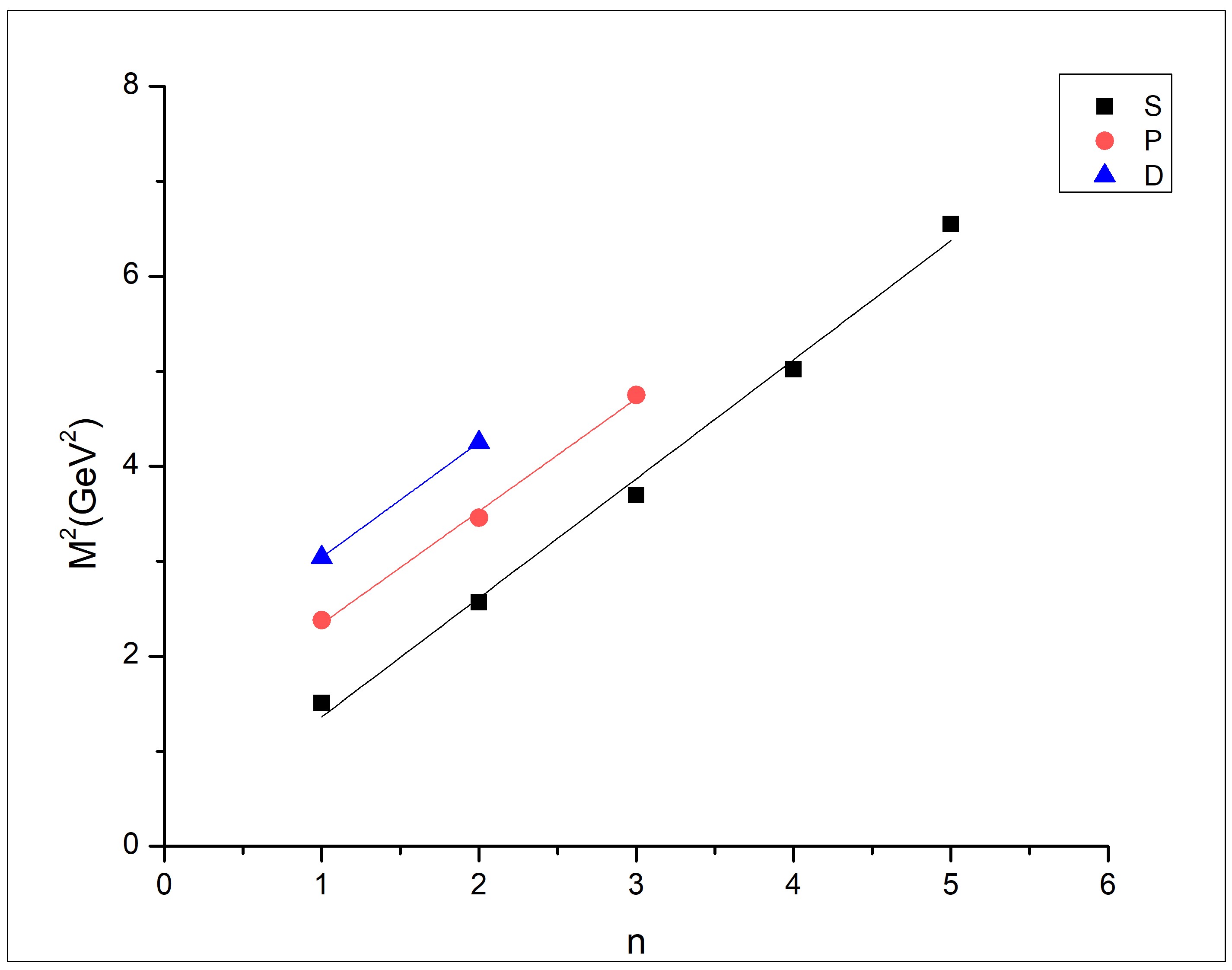}
    
  \end{minipage}
  \hfill  
  \begin{minipage}[b]{0.49\textwidth}
  \caption{Regge trajectory $J \rightarrow M^{2}$ for $\Delta^{++}$}
    \includegraphics[width=\textwidth]{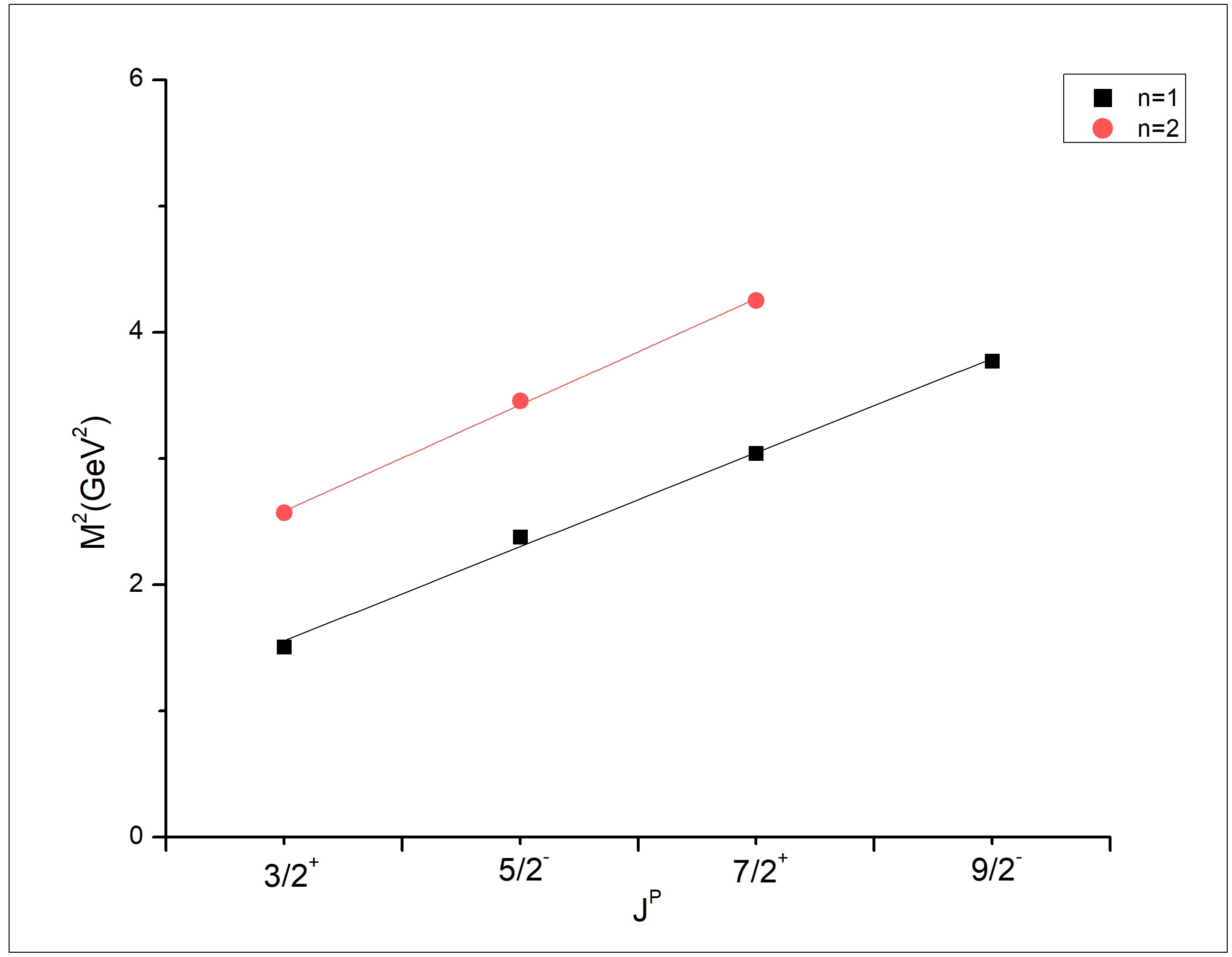}
  \end{minipage}
\end{figure}

\begin{figure}[!tbp]
  \centering
  \begin{minipage}[b]{0.49\textwidth}
  \caption{Regge trajectory $n \rightarrow M^{2}$ for $\Delta^{+}$}
    \includegraphics[width=\textwidth]{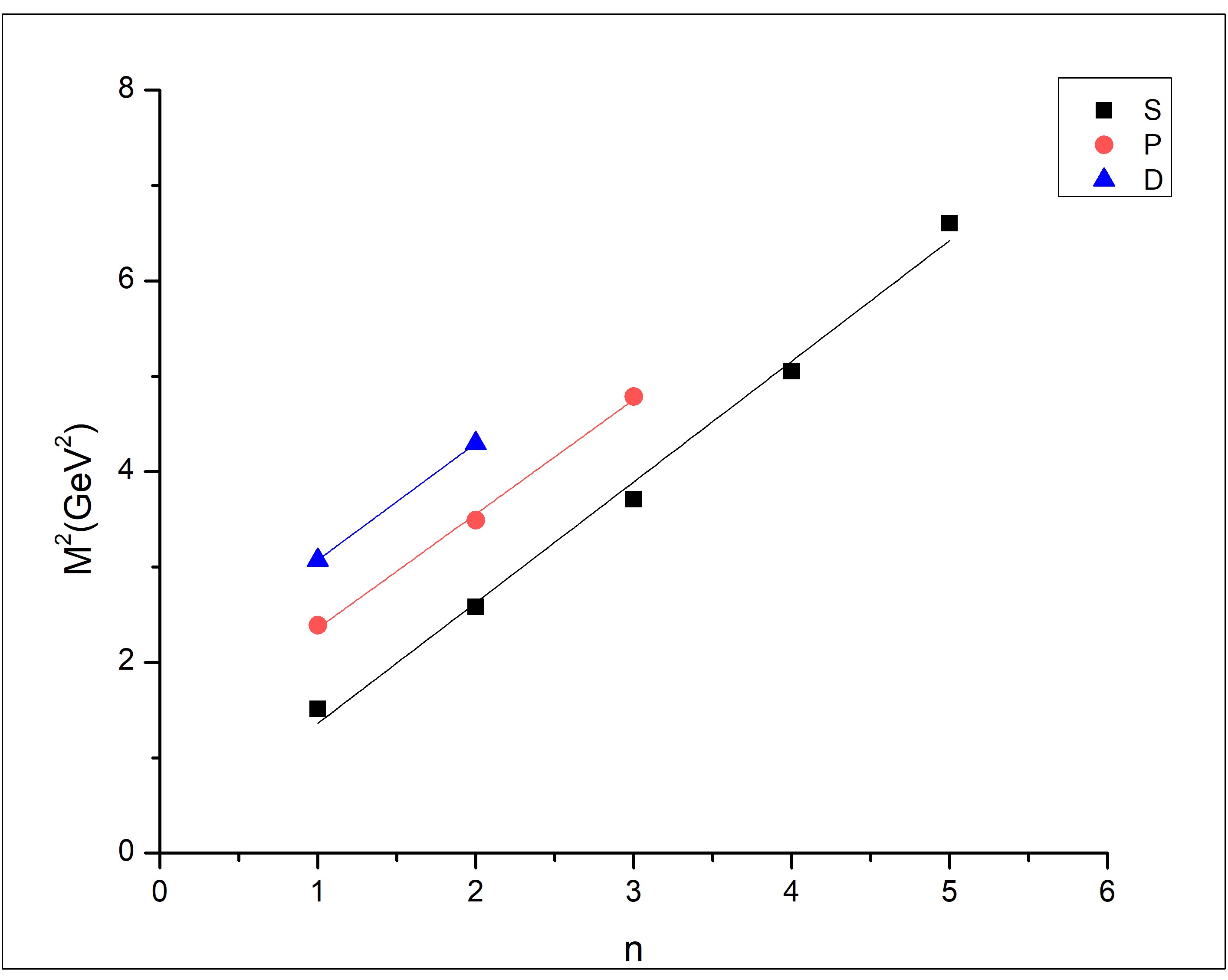}
  \end{minipage}
  \hfill  
  \begin{minipage}[b]{0.49\textwidth}
  \caption{Regge trajectory $J \rightarrow M^{2}$ for $\Delta^{+}$}
    \includegraphics[width=\textwidth]{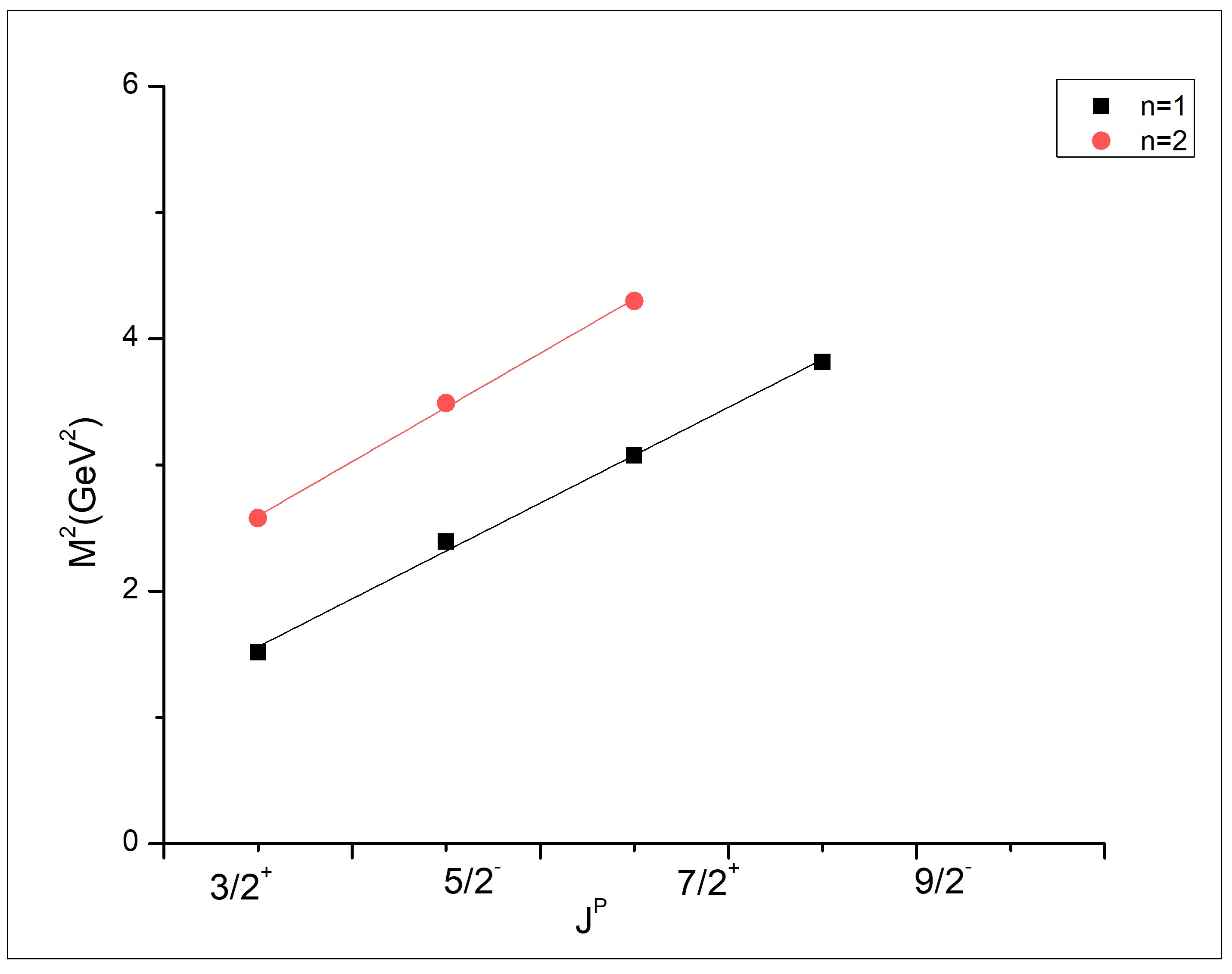}
  \end{minipage}
\end{figure}

\begin{figure}[!tbp]
  \centering
  \begin{minipage}[b]{0.49\textwidth}
  \caption{Regge trajectory $n \rightarrow M^{2}$ for $\Delta^{0}$}
    \includegraphics[width=\textwidth]{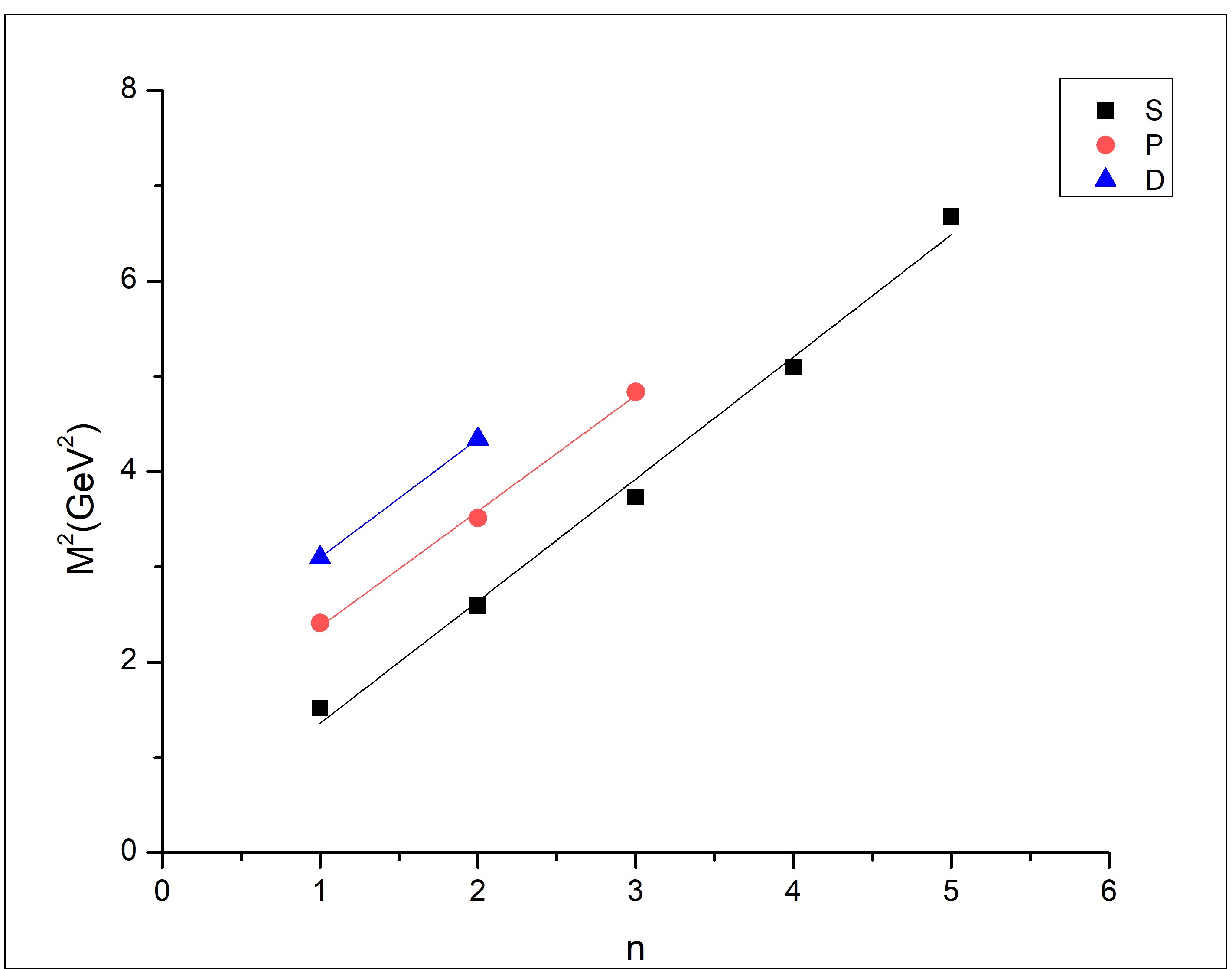}
  \end{minipage}
  \hfill  
  \begin{minipage}[b]{0.49\textwidth}
  \caption{Regge trajectory $J \rightarrow M^{2}$ for $\Delta^{0}$}
    \includegraphics[width=\textwidth]{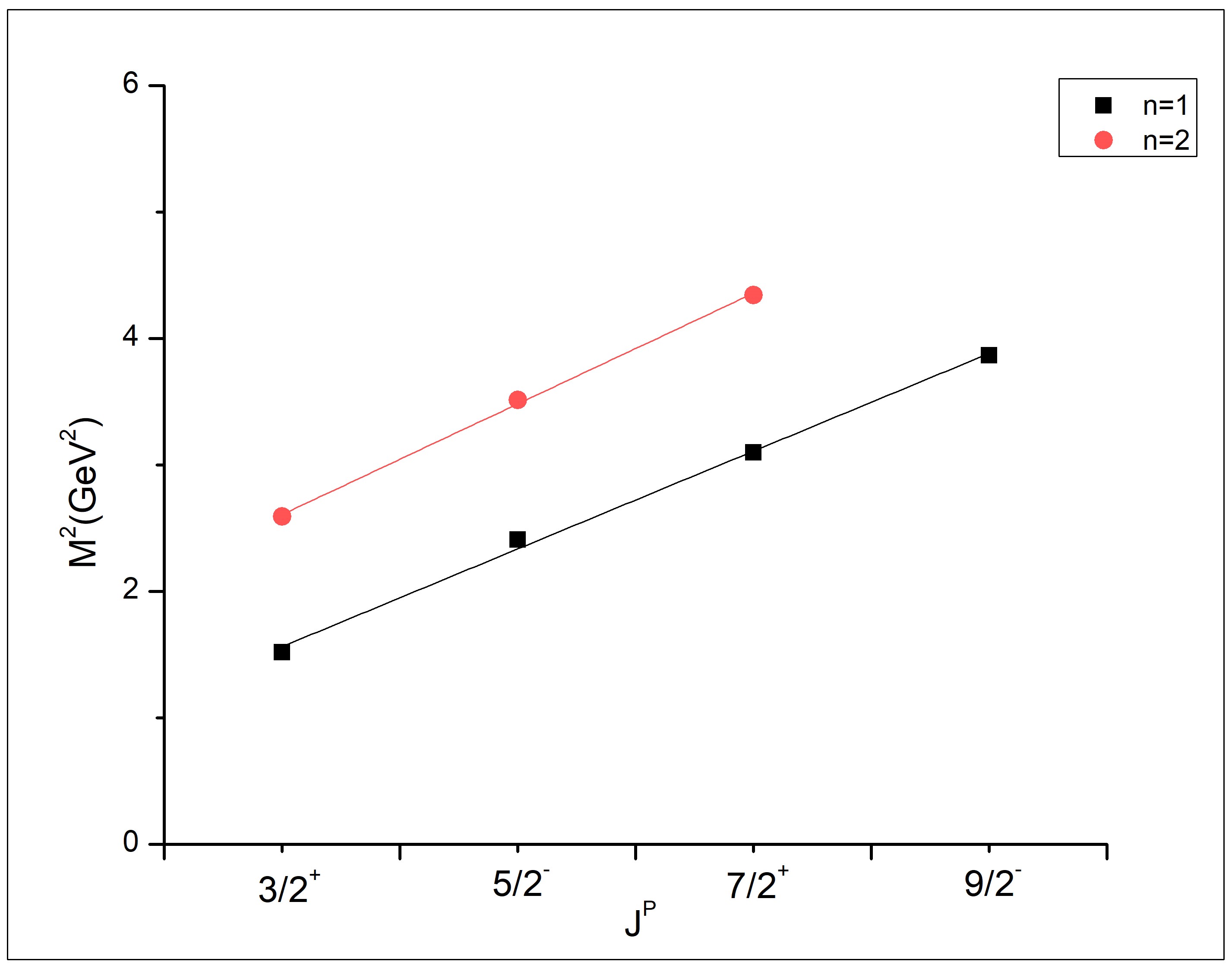}
  \end{minipage}
\end{figure}

\begin{figure}[!tbp]
  \centering
  \begin{minipage}[b]{0.49\textwidth}
  \caption{Regge trajectory $n \rightarrow M^{2}$ for $\Delta^{-}$}
    \includegraphics[width=\textwidth]{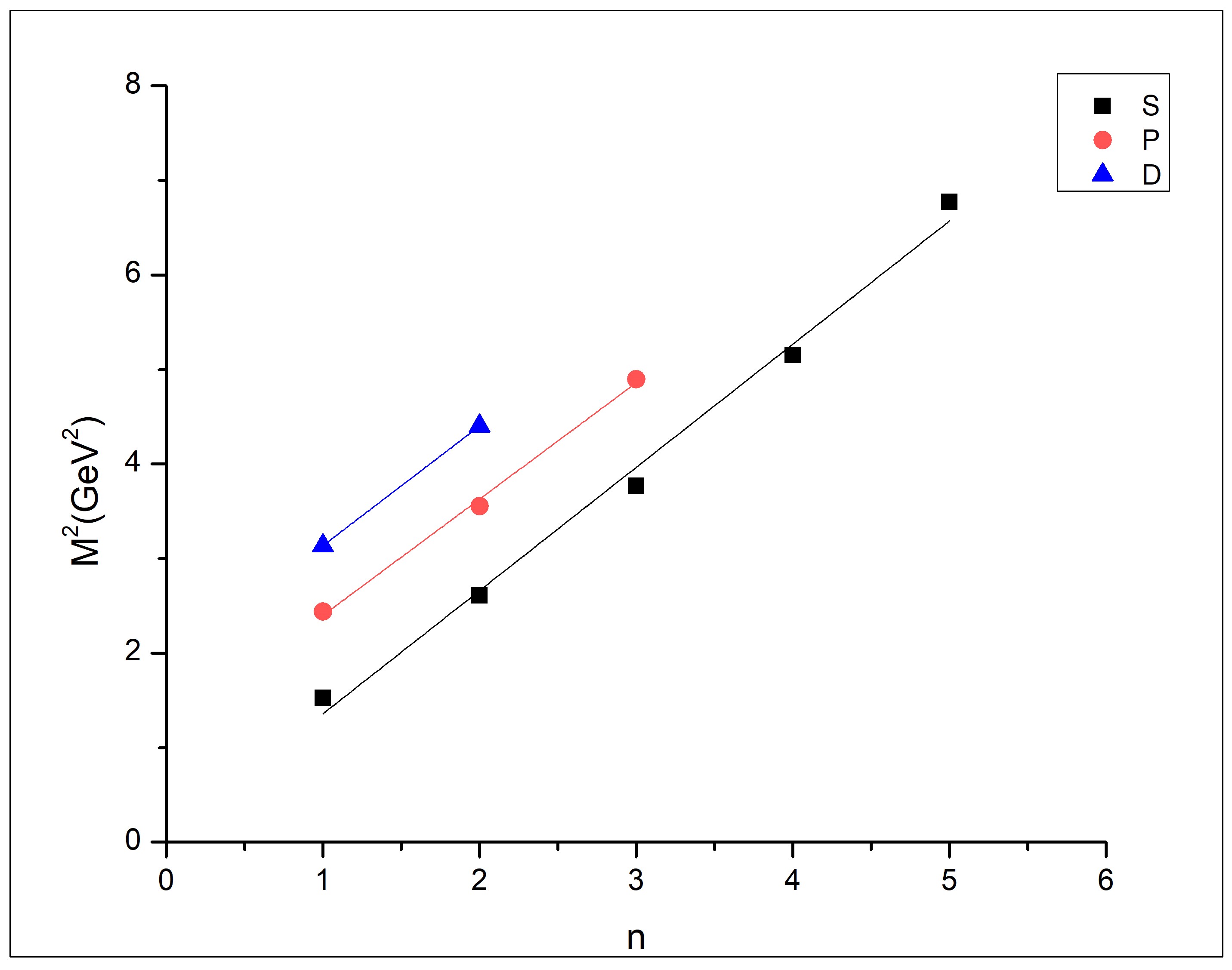}
  \end{minipage}
  \hfill  
  \begin{minipage}[b]{0.49\textwidth}
  \caption{Regge trajectory $J \rightarrow M^{2}$ for $\Delta^{-}$}
    \includegraphics[width=\textwidth]{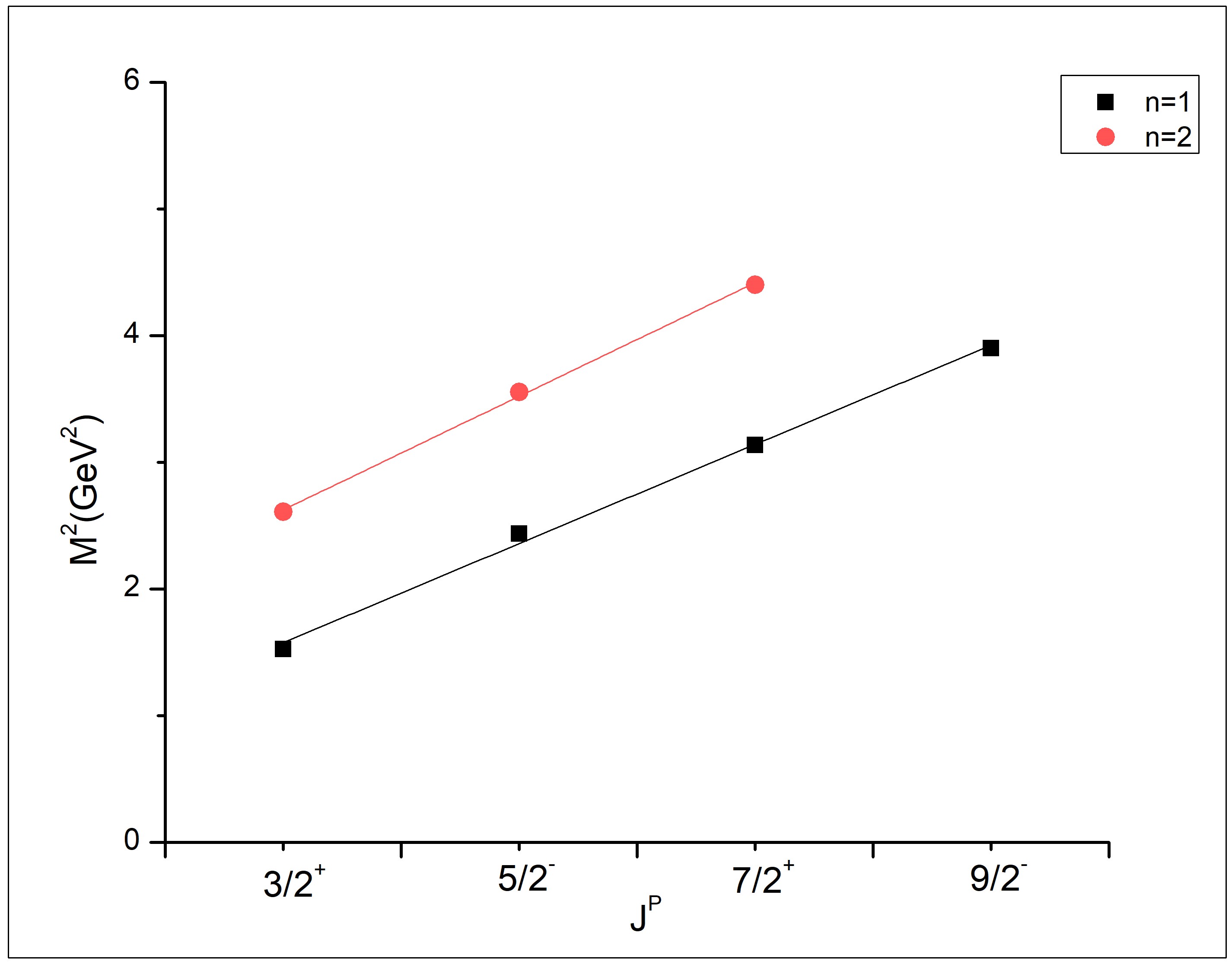}
  \end{minipage}
\end{figure}
\section{Magnetic Moment}

The electromagnetic property of baryons plays an important role in theoretical and experimental aspects.  Baryon magnetic moment assists in the understanding of internal structure of baryon. The short lifetime of decuplet baryons make it difficult to obtain their magnetic moments. It is noteworthy that baryon magnetic moments are not just the contributions of those carried by valence quarks but orbital excitations, sea quarks, relativistic effects, meson cloud effectand other effects \cite{liu}.\\\\
Baryon magnetic moments have been studied through various models over the years. H. Dahiya et. al. have employed chiral quark model incorporating sea quark polarizations and orbital angular momentum through Cheng-Li mechanism \cite{harleen3, aarti}. 


Another studies include effective quark mass and screened charge formalism \cite{dhir}, light cone QCD Sum rule \cite{aliev}, chiral perturbation theory \cite{flores}, lattice QCD \cite{lee}, statistical model \cite{kaur}, through new quark relation using mass differences and ratio \cite{karliner} and many more. \\\\
In the present article, effective quark mass which is different from that of constituent quark mass has been used to obtain the magnetic moment of all four $\Delta$ isospins. The effective quark mass is given as 
\begin{equation}
m_{q}^{eff}=m_{q}(1+\frac{\langle H \rangle}{\sum_{q}m_{q}})
\end{equation}
The magnetic moment is obtained using 
\begin{equation}
\mu_{B}= \sum_{q} \left\langle \phi_{sf} \vert \mu_{qz}\vert \phi_{sf} \right\rangle
\end{equation}
Here $\phi_{sf}$ is the spin-flavour wave function.
\begin{equation}
\mu_{qz}= \frac{e_{q}}{2m_{q}^{eff}}\sigma_{qz}
\end{equation}
The spin flavour wave function is obtained based on the symmetric configurations of u and d quarks in case of $\Delta$. The recent experimental value of $\mu(\Delta^{++})$ is 6.14 which is higher compared to our results. Also, it is evident for other states, the result is quite in accordance with other approaches. 
\setcounter{table}{1}
\begin{table}
\caption{Comparison of calculated magnetic moments with various models (All data in units of $\mu_{N}$)}
\label{tab:mmcompare}
\begin{tabular}{ccccccccccccc}
\hline
Baryon & $\mu_{cal}$ & Exp \cite{pdg} & \cite{harleen3} & \cite{harleen3} & \cite{linde}  & \cite{dhir} & \cite{dhir} & \cite{harpreet} & \cite{buchman} & \cite{flores}\\
\hline
$\Delta^{++}$ & 4.58 & 6.14 & 5.43 & 5.97 & 5.21 & 4.56 & 4.68 & 5.267 & 5.90 & 5.390 & \\
$\Delta^{+}$ & 2.34 & 2.7 & 2.72 & 2.76 & 2.45 & 2.28 & 2.36 & 2.430 & 2.90 & 2.383 & \\
$\Delta^{0}$ & 0.05 & - & 0 & -0.46 & -0.30 & 0 & -0.025 & -0.407 & - & -0.625 & \\
$\Delta^{-}$ & -2.28 & - & -2.72 & -3.68 & -3.06 & -2.28 & -2.34 & -3.245 & -2.90 & -3.632 & \\
\hline
\end{tabular}
\end{table}


%

\section{Transition Magnetic Moment}
The low-lying baryon decuplet to octet transition also play a key part in the intrinsic spin properties of baryons including deformation. As discussed in above section, many approaches have been used to obtain transition magnetic moment as well as radiative decay widths of decuplet baryons. The generalized form for transition magnetic moment is obtained by changing the spin flavour wave function for $S=\frac{3}{2}$ to $S=\frac{1}{2}$ \cite{fayyazuddin},
\begin{equation}
\mu(B_{\frac{3}{2}^{+}} \rightarrow B_{\frac{1}{2}^{+}}) = \langle B_{\frac{1}{2}^{+}}, S_{z}=\frac{1}{2}|\mu_{z}| B_{\frac{3}{2}^{+}}, S_{z}=\frac{1}{2} \rangle
\end{equation}
And the radiative decay width is obtained as follows,
\begin{equation}
\Gamma_{R}= \frac{q^{3}}{m_{p}^{2}}\frac{2}{2J+1}\frac{e^{2}}{4\pi}|\mu_{\frac{3}{2}^{+} \rightarrow \frac{1}{2}^{+}}|^{2}
\end{equation} 
where q is the photon energy, $m_{p}$ is the proton mass and J is the initial angular momentum.
In the present article, we study the transition of $\Delta^{+} \rightarrow P\gamma$ and $\Delta^{0} \rightarrow N\gamma$.
The expressions for both these channels shall be as 
\begin{equation}
\frac{2\sqrt{2}}{3}(\mu_{u}^{eff}-\mu_{d}^{eff}) \quad \& \quad \frac{2\sqrt{2}}{3}(\mu_{d}^{eff}-\mu_{u}^{eff})
\end{equation}
Here, the effective mass is a geometric mean of those for spin $\frac{1}{2}$ and $\frac{3}{2}$ as
\begin{equation}
m_{i}^{eff}=\sqrt{{m_{i}^{3/2}}{m_{i}^{1/2}}}
\end{equation} 
The separate masses for P and N are taken from our previous article \cite{epjweb} as 938 and 948 MeV respectively. $M_{\Delta^{+}} = 1230 $ MeV and $M_{\Delta^{0}} = 1232 $ MeV
\begin{table}
\centering
\caption{Transition magnetic moment (in $\mu_{N}$)}
\label{tab:radiative}
\begin{tabular}{cccccc}

\hline
Decay &  Transition moment(in $\mu_{N}$) & \cite{harleen18} & \cite{dhir} & \cite{ramalho} & \cite{hong}\\
\hline
$\Delta^{+} \rightarrow P\gamma$ & 2.47 & 3.87 & 2.63 & 3.32 & -2.76 \\
$\Delta^{0} \rightarrow N\gamma$ & -2.48 & \\
\hline
\end{tabular}
\end{table}
The radiative decay width obtained using equation (17) is 0.63 MeV and 0.59 MeV for $\Delta^{+}$ and $\Delta^{0}$ respectively. Our results for fraction of 0.54\% and 0.5\% are well in accordance with the experimental range of 0.55-0.65\%. 

\section{Conclusion}
The idea of separately exploring the isospin states of $\Delta$ baryon has been implemented in the present work which is a modification to earlier calculated masses. The constituent quark masses for u and d quarks have been treated differently to obtain the radial and orbital excited state masses using the non-relativistic hypercetral Constituent Quark Model (hCQM). The potential incorporated consists of linear confining term, spin-dependent terms and first order correction terms. \\\\
The results have been compared with those of Particle Data Group (PDG) and with other approaches discussed above. The comparison have shown that low-lying states are well in accordance with experimental range. However, the higher $J^{P}$ value states for a given principle quantum number under-predicts as compared to PDG. The results are expected to aid in future experiments to determine missing resonances and in various decay channels.\\\\
The Regge trajectories have been plotted for (n,$M^{2}$) and  (J,$M^{2}$) which are following the linear nature. This allows us to identify a given state for its spin-parity assignment on the Regge line. The magnetic moment for $\Delta^{++}$ and $\Delta^{+}$ are only known through PDG and the results vary within 1$\mu_{N}$ similarly for all other compared models. The magnetic moment for electromagnetic transition of channels $\Delta^{+} \rightarrow P\gamma$ and $\Delta^{0} \rightarrow N\gamma$ are well in accordance with other approaches. The radiative decay width giving 0.54\% and 0.5\% also fall around 0.55-0.65\% range. \\\\
Thus, a non-relativistic CQM with linear confinement potential has provided us with a number of resonance masses with all possible spin-parity assignments. Differentiating the isospin resonance masses shall be important to study strong, weak and electromagnetic decay channels. This work is expected to support the upcoming experimental facilities at PANDA \cite{panda1,panda2, panda3, panda4}.

\section{Acknowledgement}
The authors are thankful to the organizers of 10th International Conference on New Frontiers in Physics (ICNFP 2021) for providing with an opportunity to present our work. 
Also, Ms. Chandni Menapara would like to acknowledge the support from the Department of Science and Technology (DST) under INSPIRE-FELLOWSHIP scheme for pursuing this work.\\

\end{document}